\documentstyle[12pt]{article}
\begin{document}
\baselineskip=22pt plus 0.2pt minus 0.2pt
\lineskip=22pt plus 0.2pt minus 0.2pt
\hspace*{12cm}LAEFF 95-25
\vspace*{3cm}
\begin{center}
 \Large
From Euclidean to Lorentzian General\\
Relativity: The Real Way\\

\vspace*{0.35in}

\large

J.\ Fernando\ Barbero\ G. $^{\ast}$
\vspace*{0.25in}

\normalsize

$^{\ast}$Laboratorio de Astrof\'{\i}sica Espacial y F\'{\i}sica
Fundamental, INTA,\\
P.O. Box 50727,\\
28080 Madrid, SPAIN\\

\vspace{.5in}
June 5, 1995\\
\vspace{.5in}
ABSTRACT
\end{center}
We study in this paper a new approach to the problem of relating
solutions to the Einstein field equations with Riemannian and
Lorentzian signatures. The procedure can be thought of as a ``real
Wick rotation". We give a modified action for general relativity,
depending on two real parameters, that can be used to control the
signature of the solutions to the field equations. We show how
this procedure works for the Schwarzschild metric and discuss some
possible applications of the formalism in the context of signature
change, the problem of time, black hole thermodynamics...


\pagebreak

\setcounter{page}{1}

\section{Introduction} 

This paper is devoted to the study of the relationship between
Riemannian (referred to in the following as ``Euclidean") and
Lorentzian signature solutions to the Einstein field equations
(EFE's). 

The Schwarzschild metric is a good example for discussing the
importance of having the possibility of relating Euclidean and
Lorentzian metrics, and illuminates why the usual approach fails.
For Lorentzian signatures it is
\begin{equation}
ds^2=-\left(1-\frac{2M}{r}\right)dt^2+
\left(1-\frac{2M}{r}\right)^{-1}
dr^2+r^2\left(d\theta^2 + \sin^2\theta\;d\phi^2\right)
\label{1}
\end{equation}
In this case there is a simple way to obtain a Euclidean solution to
the EFE's: introduce a complex time variable $\tau=it$, rewrite
(\ref{1}) in the new coordinate system $(\tau,\;r,\;\theta,\; \phi)$,
and take advantage of the fact that the components of the metric do
not have any explicit time dependence. We get then
\begin{equation}
ds^2_E=\left(1-\frac{2M}{r}\right)d\tau^2+
\left(1-\frac{2M}{r}\right)^{-1} dr^2+
r^2\left(d\theta^2 + \sin^2\theta\;d\phi^2\right)
\label{2}
\end{equation}
which has Euclidean signature. The properties of (\ref{2}) are useful
in the study of the thermodynamics of a Schwarzschild black hole (see
for example \cite{Wald84}\cite{Wald95}); and in particular to
understand the thermal nature of the Hawking radiation. 

The strategy of analytic continuation used in the previous example
consists, more precisely, in trying to find a four-dimensional complex
manifold endowed with a complex metric, solution to the EFE's, such
that different real sections (real four-dimensional manifolds) exist,
some of them with Euclidean and others with Lorentzian
signature. The problem with this approach is that it can
only be made to work in some very special examples, 
static space-times, for which it is always possible to find
coordinates such that the $g_{0i}$ components of the metric are zero
and the rest of them time-independent. It is known that this method
cannot be used in general and hence an alternative must be found.

Another context in which the use of Euclidean metrics (and even
Euclidean metrics that evolve into Lorentzian ones) is useful is the
Euclidean path integral approach to quantum cosmology and quantum
gravity \cite{HH}. Some of the reasons for this are technical. In many
cases, the Euclidean action is exponentially damped instead of being
oscillatory as its Lorentzian counterpart thus improving the
convergence properties of the path integral.  Lorentzian propagators
and related objects are obtained as analytic continuations of the
Euclidean ones. From the point of view of quantum cosmology some of
the most appealing models for the very early universe describe its
origin as a quantum tunneling from a Euclidean regime to a Lorentzian
one, so the general problem of understanding how to ``connect"
Euclidean and Lorentzian space-times, both at the quantum and the
classical level, seems to be an important one.

The main result of this paper is a prescription to
find families of metrics parametri\-zed by two real numbers $\alpha,
\beta$ satisfying the following properties:

i) All of them are solutions to the vacuum EFE's (except for a zero
measure set of values of  $\alpha,\beta$).

ii) Some members of the family have Euclidean signature whereas the
others have Lorentzian signature.

The lay-out of the paper is the following. After this introduction we
discuss in section II a modified action principle for general
relativity that describes both Euclidean and Lorentzian vacuum
solutions for general relativity in terms of an auxiliary metric
field. We study this by considering the Hamiltonian formulation of the
theory. In section III we discuss the interpretation of the
constraints derived from the new action. Section IV concentrates on
the study of the new field equations, and especially on how to
construct solutions to the EFE's from them. We discuss in section
V a particular, and physically relevant example: Schwarzschild
space-times; and show that we recover the usual Euclidean and
Lorentzian solutions (\ref{1}) and (\ref{2}). We end the paper in
section VI with some comments, conclusions and suggestions for future
work within this new approach.

\section{The Modified Action For General Relativity}

The main point of this paper is the discussion of a new action
principle for general relativity. In what follows we will use
the Hamiltonian formulation of the theory extensively. As is well
known the ADM \cite{ADM} formulations for Euclidean and Lorentzian
general relativity share the same phase space (coordinatized by
3-metrics $q_{ab}$ and their canonically conjugate momenta
$\tilde{p}^{ab}$) and the same symplectic structure (i.e. Poisson
brackets). The difference between both theories reduces ``only" to
the relative sign between the potential and kinetic terms in the
Hamiltonian constraint. As it is important to be able to track the
minus signs, and for the benefit of the reader, we give some details
about the derivation of the ADM Hamiltonian; we will also use some of
this results to study the new action. For most of this part we
follow\cite{Wald84}.

The notation used in the paper is the following. Tangent space
indices are denoted as $a,\;b,\;c\ldots$ going from 0 to 3. We will
not use three-dimensional indices; tensor
fields in three dimensions will be naturally described by their
projection properties onto three-dimensional manifolds. Let us
consider a four-dimensional manifold
${\cal M}=\Sigma\times {\rm l\!R}$ where the three-dimensional
$\Sigma$ is either compact without a boundary or, if not compact, the
fall-off of the fields is taken such that the possible
surface terms that appear do not give any contribution. We introduce
in ${\cal M}$ a metric $g_{ab}$ with Euclidean signature (++++).
$\cal{M}$
can be foliated by three-dimensional surfaces defined by the constant 
value of a certain scalar function t. We take the foliation in the
``non-compact direction" corresponding to ${\rm l\!R}$. Given this
foliation we can write the gradient 1-form\footnote{$\partial_{a}$ is
any torsion-free derivative in ${\cal M}$.}
\begin{equation}
dt=(\partial_{a}t)dx^{a}
\label{3}
\end{equation}
With the help of the inverse 4-metric $g^{ab}$ we can define a unit,
future directed, normal to the foliation
\begin{equation}
n_{a}=\frac{\partial_{a}t}{(g^{bc}\partial_{b}t \partial_c
t)^{1/2}},\;\;n^{a}\equiv g^{ab}n_{b},\;\;n_{a}n^{a}=1
\label{4}
\end{equation}
The 3-metric $q_{ab}$ (first fundamental form) induced in the sheets
of the foliation by the four-dimensional $g_{ab}$ is
\begin{equation}
\begin{array}{l}
q_{ab}\equiv g_{ab}-n_a n_b\\
q^{ab}\equiv g^{ab}-n^a n^b\\
q_{a}^{\;\;b}\equiv\delta_{a}^{\;\;b}-n_{a}n^{b}=q_{ac}q^{cb}=
q_{ac}g^{bc}
\end{array}
\label{5}
\end{equation}
where we have made use of the fact that $q_{ab}n^{b}=0$. Notice that
$q_{a}^{\;\;b}$ is a projector onto the three-dimensional sheets of
the foliation. We give now a mapping that relates points in different
sheets by using a congruence of curves filling the manifold ${\cal M}$
and never tangent to them. Two points in different sheets are
``connected" if they are on the same curve of the congruence. In a
rather loose sense the foliation allows us to say if two points of
${\cal M}$ are ``simultaneous" and the congruence if they are ``at the
same space point" (see \cite{KU81}). We parametrize the curves in
the congruence with the help of t. If one such curve is given by the
functions $x^{a}(t)$ we have
\begin{equation}
dt=(\partial_{a}t) dx^{a}\Rightarrow 1=\partial_{a}t
\frac{dx^{a}}{dt}\equiv t^{a}\partial_at
\label{6}
\end{equation}
where $t^a$ is the tangent vector to the congruence. We define
``partial time derivatives" as Lie derivatives along the direction
given by $t^a$. Notice that this is consistent because (\ref{6})
implies
\begin{equation}
\dot{t}\equiv {\cal L}_{\vec{t}} \;t=t^a\partial_a t=1
\label{7}
\end{equation}
Another way to see this is by building a coordinate system in ${\cal
M}$ using t as the $x^0$ coordinate, giving coordinates
$(x^1,x^2,x^3)$ to one sheet of the foliation and Lie-dragging 
them to the rest of ${\cal M}$ by using the congruence of curves. It
is obvious that by proceeding in this manner, $\partial_t$ is the
same as
${\cal L}_{\vec{t}}$.

At each point P of ${\cal M}$ the vector field $t^a$ can be uniquely
written as a vector tangent to the foliation passing through P and a
vector normal to the foliation at P.
\begin{equation}
t^a=N n^a+ N^a
\label{8}
\end{equation}
$N$  and  $N^a$ are known as the lapse and the shift respectively, and
satisfy the following properties
\begin{equation}
N^a n_a=0,\;\;N=t^a n_a,\;\;N=\frac{1}{n^a \partial_a
t}=\frac{1}{(g^{ab}\partial_a t\partial_b t)^{1/2}}
\label{9}
\end{equation}
Without loss of generality we can take $N>0$.
The extrinsic curvature of the foliation $K_{ab}$ (second fundamental
form) is defined by
\begin{equation}
K_{ab}=q_{a}^{\;\;c} q_{b}^{\;\;d}\nabla_{c}n_{d}=q_{a}^{\;\;c}
\nabla_{c}n_b
\label{10}
\end{equation}
and is a symmetric tensor that ``lives" on the sheets of the
foliation, i.e. $K_{ab}n^a=0$, $q_{a}^{\;\;b}K_{bc}=K_{ac}$. In the
previous expressions $\nabla_{a}$ denotes the (covariant) derivative
operator compatible with the 4-metric $g_{ab}$. It is possible to 
define a three-dimensional covariant derivative on objects tangent to
the foliation $S_{a_1...}^{\;\;\;\;b_1...}$ as
\begin{equation}
{\cal D}_{a} S_{a_1...}^{\;\;\;\;b_1...}=
q_{a}^{\;\;f}q_{a_1}^{\;\;f_1}
\ldots q_{g_1}^{\;\;b_1}\nabla_f S_{f_1...}^{\;\;\;\;g_1...}
\label{11}
\end{equation}
It is straightforward to show that this is the unique operator
compatible with $q_{ab}$. With the help of ${\cal D}_{a}$ we can build
the intrinsic curvature of the sheets of the foliation and obtain the
Gauss-Codazzi equations
\begin{eqnarray}
& & ^{3}R_{abc}^{\;\;\;\;\;\;d}=q_{a}^{\;\;e}
q_{b}^{\;\;f}q_{c}^{\;\;g}
q_{h}^{\;\;d}\; {}^{4}
\!R_{efg}^{\;\;\;\;\;h}+K_{ac}K_{b}^{\;\;d}-K_{bc}K_{a}^{\;\;d}
\label{12}\\
& & {\cal D}_a K^{a}_{\;\;b}-{\cal D}_{b}K=
{}^4\!R_{fe}n^{f}q^{e}_{\;\;b}
\label{13}
\end{eqnarray}
Contracting (\ref{12}) with $q^{ac}q^{bd}$ we obtain
\begin{equation}
2G_{ab} n^{a}n^{b}=-^3\!R+K^2-K_{ab} K^{ab}
\label{14}
\end{equation}
where $G_{ab}\equiv R_{ab}-\frac{1}{2}g_{ab} R$ is the
four-dimensional
Einstein tensor and $K\equiv q^{ab}K_{ab}$. Before we introduce the
new action we give an additional identity that we will use in order to
obtain the momenta canonically conjugate to $q_{ab}$ in the
Hamiltonian formulation
\begin{equation}
{\dot q}_{ab}\equiv {\cal L}_{\vec{t}}\;q_{ab}=2NK_{ab}+2{\cal
D}_{(a}N_{b)}
\label{15}
\end{equation}
Let us consider now the following action defined in ${\cal M}$ as
\begin{equation}
S[g_{ab};\alpha,\beta]=\int_{{\cal M}} d^4x\sqrt{g}
\left[\alpha G^{ab}+\beta R^{ab}\right]\eta_a \eta_b
\label{16}
\end{equation}
where $\alpha$ and $\beta$ are constant real parameters and
\begin{equation}
\eta_a=\frac{\partial_a\Phi}
{(g^{bc}\partial_b\Phi\partial_c\Phi)^{1/2}}
\label{17}
\end{equation}
$\Phi$ is a {\it fixed} scalar function monotonically growing in the
${\rm l\!R}$ direction that can be interpreted as an external time
variable\footnote{The possibility of taking $\Phi$ as a dynamical
field will be explored below.}. The previous action is non-covariant 
in the sense that there
is a fixed field related to the structure of space-time that appears
in the action and, hence, in the field equations (see, for
example the discussion in \cite{Wald84}). Before we continue, and with
the object of making the reader less suspicious with respect to
(\ref{16}), we make here the following comment (a longer discussion
will appear at the end of the paper). In the following we are going to
perform a particular Legendre transform on (\ref{16}). In order to do
that we first write (\ref{16}) as the time integral (or rather the
integral in $\Phi$) of some Lagrangian
whose variables are defined with the help of a particular foliation of
${\cal M}$: that given by the level surfaces of the scalar function
$\Phi$. After that we perform a Legendre transform to obtain the
Hamiltonian and derive the constraints. The dynamics defined by them
can be shown to be that of
general relativity (for most values of $\alpha$ and $\beta$). We will
see that we can choose the space-time signature simply by changing the
value of these parameters. Of course foliations other than the
one given by $\Phi$ can be used. In that case the Legendre transform
is more complicated to perform but the dynamics is the same as the one
obtained by using the particular foliation given by $\Phi$ because
they are derived by performing Legendre transforms on the same action
functional.

By using (\ref{14}) we get
\begin{equation}
\alpha\int_{{\cal M}} d^4x\sqrt{g}\;G^{ab}\eta_a\eta_b=
\frac{\alpha}{2}\int_{{\cal M}} d^4x\sqrt{g}\left[-{}^{3}
\!R+K^2-K_{ab}K^{ab}\right]
\label{18}
\end{equation}
where all the three-dimensional quantities refer to the foliation
defined by $\Phi$. The second term in (\ref{16}) can be written as
\begin{eqnarray}
& &\beta\int_{{\cal M}} d^4x\sqrt{g}R^{ab}\eta_a\eta_b=\beta
\int_{{\cal M}} d^4x\sqrt{g}\left\{\nabla_{a}\left[
\eta^b\nabla_b \eta^a
\right]- \nabla_b\left[\eta^b\nabla_a
\eta^a\right]+\right.\hspace{2cm}
\label{19}\\
& & \hspace{7.5cm}+\left.(\nabla_b \eta^b)(\nabla_a \eta^a)-
(\nabla_a \eta^b)(\nabla_b \eta^a)\right\}\nonumber
\end{eqnarray}
The first two terms are surface terms. If $\Sigma $ is compact they
are not present. If $\Sigma$ is non-compact they are not zero but do
not give any contribution when varying the action because the
variations of the fields are zero at the boundaries\footnote{Surface
terms are important for the consistency of the Hamiltonian 
formulation; the usual Einstein-Hilbert action must be modified with
the addition of surface integrals at the boundaries. None of these
will change the arguments presented here.}. With this remarks in mind
we see that
\begin{equation}
\beta\int_{{\cal M}} d^4x\sqrt{g}\;R^{ab}\eta_a\eta_b=\beta
\int_{{\cal M}} d^4x\sqrt{g}\left[K^2-K_{ab}K^{ab}\right]
\label{20}
\end{equation}
Combining (\ref{18}) and (\ref{20}) we finally get
\begin{equation}
S[g_{ab};\alpha,\beta]=\int_{{\cal M}}
d^4x\sqrt{g}\left[-\frac{\alpha}{2}\;{}^3\!R
+(\frac{\alpha}{2}+\beta)\left(K^2-K_{ab}K^{ab}\right)\right]
\label{21}
\end{equation}
The measure in the previous integral can be written as
\begin{equation}
d^4x\sqrt{g}=d\Phi\;d^3x N\sqrt{q}
\label{22}
\end{equation}
Notice that $d^3x \sqrt{q}$ is the measure in the
three-dimensional slices
of the foliation defined by the induced metric $q_{ab}$. We have then
\begin{equation}
S[g_{ab};\alpha,\beta]=\int_{\rm l\!R}d\Phi\int_{\Sigma_\Phi}
d^3x\sqrt{q}N\left[-\frac{\alpha}{2}\;{}^3\!R
+(\frac{\alpha}{2}+\beta)\left(K^2-K_{ab}K^{ab}\right)\right]
\label{23}
\end{equation}
The Lagrangian $L$ can be read off directly from the previous formula. 
In order to obtain the corresponding Hamiltonian we perform a Legendre
transform in the usual way. We first define the momenta canonically
conjugate to the configuration variable $q_{ab}$.
\begin{equation}
\tilde{p}^{ab}(x)=\frac{\delta L}{\delta\dot{q}_{ab}(x)}=
\int_{\Sigma_{\Phi}}d^3y \frac{\delta L}{\delta K_{cd}(y)}
\frac{\delta K_{cd}(y)}{\delta \dot{q}_{ab}(x)}=
\sqrt{q}(\frac{\alpha}{2}+\beta)\left(Kq_{ab}-K_{ab}\right)
\label{25}
\end{equation}
which implies
\begin{equation}
K_{ab}=\frac{2}{\alpha+2\beta}\frac{1}{\sqrt{q}}\left[\frac{1}{2}
q_{ab}\tilde{p}-\tilde{p}_{ab}\right],\;\;\alpha+2\beta\neq
0,\;\;\tilde{p}\equiv q^{ab}\tilde{p}_{ab}
\label{26}
\end{equation}
When $\alpha+2\beta=0$ we find the primary constraint
$\tilde{p}^{ab}=0$ and the theory is very different from the ones with
$\alpha+2\beta\neq 0$. We will not discuss this here although
understanding its meaning may be relevant for the study of signature
change. The Hamiltonian is now

\begin{equation}
H=\int_{\Sigma_\Phi}d^{3}x\left\{N\left[\frac{\alpha}{2}\sqrt{q}\;{}^{3}
\!R+\frac{2}{\alpha+2\beta}\frac{1}{\sqrt{q}}\left(\frac{1}{2}
\tilde{p}^2-\tilde{p}^{ab}\tilde{p}_{ab}\right)\right]
-2N_{a}{\cal D}_b\tilde{p}^{ab}\right\}
\label{30}
\end{equation}
and the constraints are
\begin{eqnarray}
& & \frac{\alpha}{2}\sqrt{q}\;{}^{3}\!R+
\frac{2}{\alpha+2\beta}\frac{1}{\sqrt{q}}\left(\frac{1}{2}\tilde{p}^2-
\tilde{p}^{ab}\tilde{p}_{ab}\right)=0\label{31}\\
& & {\cal D}_b\tilde{p}^{ab}=0\label{32}
\end{eqnarray}

\section{Discussion and Interpretation of the New Constraints}

As it is well known the ADM constraints for general relativity in the
Euclidean and Lorentzian formulations differ only in the relative
signs between the kinetic and potential terms, that is
\begin{eqnarray}
& & {\rm EUCLIDEAN} \hspace{1cm} -\sqrt{q}\;{}^{3}\!R+\frac{1}{\sqrt{q}}
\left(\frac{1}{2}\tilde{p}^2-\tilde{p}_{ab}\tilde{p}^{ab}\right)=0
\label{33}\\
& & {\rm LORENTZIAN} \hspace{.7cm}+\sqrt{q}\;{}^{3}\!R+
\frac{1}{\sqrt{q}}
\left(\frac{1}{2}\tilde{p}^2-\tilde{p}_{ab}\tilde{p}^{ab}\right)=0
\label{34}
\end{eqnarray}
We can see that we can get either (\ref{33}) or (\ref{34}) from
(\ref{31}) by a suitable choice of the parameters $\alpha$ and $\beta$.
If we take $\alpha=-2$, $\beta=+2$ we get (\ref{33}) whereas
$\alpha=+2$, $\beta=0$ gives (\ref{34}). We arrive at the conclusion
then, that the action (\ref{16}) describes both Euclidean and
Lorentzian general relativity depending on the values chosen for the
parameters. A question arises now: What is the meaning of the action
for values of $\alpha$ and  $\beta$ other than the ones considered
above? In particular we would like to know if the constraints
(\ref{31}) and (\ref{32}) are first class for arbitrary values of the
parameters. To answer these questions let us consider the following
canonical transformation with $k\in {\rm l\!R}$ and $k>0$
\begin{equation}
\begin{array}{l}
q_{ab}\rightarrow k q_{ab}\\
{}\\
\tilde{p}^{ab}\rightarrow k^{-1}\tilde{p}^{ab}
\end{array}
\label{35}
\end{equation}
We impose $k>0$ in order to keep the signature of the 3-metric
$q_{ab}$ positive. Under this transformation we have
\begin{equation}
\begin{array}{l}
{}^{3}\!R\rightarrow  k^{-1}\; {}^{3}\!R\\
{}\\
\sqrt{q}\rightarrow k^{3/2}\sqrt{q}
\end{array}
\label{36}
\end{equation}
so (\ref{31}) and (\ref{32}) will become
\begin{eqnarray}
& & \frac{\alpha}{2}k^{1/2}\sqrt{q}\;{}^{3}\!R+
\frac{2}{(\alpha+2\beta)k^{3/2}}\frac{1}{\sqrt{q}}\left(\frac{1}{2}
\tilde{p}^2-\tilde{p}^{ab}\tilde{p}_{ab}\right)=0\label{37}\\
& & \frac{1}{k}{\cal D}_b\tilde{p}^{ab}=0\label{37b}
\end{eqnarray}
As we can see we cannot change the relative sign between the kinetic
and potential terms in the Hamiltonian constraint, but we can choose
$k$ in such a way that we get a Hamiltonian constraint proportional
either to (\ref{33}) or (\ref{34}) (see Fig. 1).
 
This proves that (\ref{31})
and (\ref{32}) are first class constraints for arbitrary values of the
parameters with the exception of $\alpha+2\beta=0$ (for which they are
not defined). Notice that the
constraints are also first class for $\alpha=0$ although the theory is
neither Euclidean nor Lorentzian general relativity. It is important
to point out, also, that the canonical transformation introduced
cannot be thought of as a coordinate transformation. First of
all because the coordinates are really kept fixed and second,
and more importantly,
because there is no coordinate transformation that can simultaneously
change the components of the covariant tensor $q_{ab}$ and the
contravariant tensor density
$\tilde{p}^{ab}$ as in (\ref{35}).

The evolution of initial data satisfying the constraints (\ref{31})
and (\ref{32}) is given by the Hamiltonian (\ref{30}). The evolution
equations for $q_{ab}$ and $\tilde{p}^{ab}$ are
$$\dot{q}_{ab}=\frac{2N}{\alpha+2\beta}\frac{1}{\sqrt{q}}
\left(\tilde{p}
q_{ab}-2\tilde{p}_{ab}\right)+2q_{c(a}{\cal D}_{b)}N^{c}$$
\begin{eqnarray}
& &\dot{\tilde{p}}^{ab}=\frac{\alpha}{2}\sqrt{q}N
\left[R^{ab}-\frac{1}{2}
q^{ab}\right]-\frac{N}{\alpha+2\beta}\frac{1}{\sqrt{q}}q^{ab}\left[
\tilde{p}^{cd}\tilde{p}_{cd}-\frac{1}{2}\tilde{p}^{2}\right]+
\label{42}\\
& &\hspace{1cm}+\frac{2N}{\alpha+2\beta}\frac{1}{\sqrt{q}}\left[2
\tilde{p}^{a}_{\;\;c}\tilde{p}^{cb}-\tilde{p}\tilde{p}^{ab}\right]+
\frac{\alpha}{2}\sqrt{q}\left[q^{ab}\Box N-{\cal D}^{a}{\cal D}^{b}
N\right]+{\cal L}_{\vec{N}}\tilde{p}^{ab}\nonumber
\end{eqnarray}
In the spirit of the canonical transformation (\ref{35}) we introduce
now
\begin{equation}
\begin{array}{l}
h_{ab}\equiv \kappa q_{ab}\\
{}\\
\tilde{\pi}^{ab}\equiv\kappa^{-1}\tilde{p}^{ab}
\end{array}
\label{43}
\end{equation}
The evolution equations for $h_{ab}$ and $\tilde{\pi}^{ab}$ are then
\begin{eqnarray}
& &\dot{h}_{ab}=\frac{2N\kappa^{3/2}}{\alpha+2\beta}\frac{1}{\sqrt{h}}
\left(\tilde{\pi}
h_{ab}-2\tilde{\pi}_{ab}\right)+2h_{c(a}{\cal D}_{b)}N^{c}\nonumber\\
& &\dot{\tilde{\pi}}^{ab}=\frac{\alpha}{2}\sqrt{h}\kappa^{-1/2} N
\left[R^{ab}(h)-\frac{1}{2}R(h)h^{ab}\right]-
\frac{N\kappa^{3/2}}{\alpha+2\beta}\frac{1}{\sqrt{h}}h^{ab}
\left[ \tilde{\pi}^{cd}\tilde{\pi}_{cd}-\frac{1}{2}\tilde{\pi}^{2}
\right]+\label{44}\\
& &\hspace{1cm}+\frac{2N\kappa^{3/2}}{\alpha+2\beta}\frac{1}{\sqrt{h}}
\left[2\tilde{\pi}^{a}_{\;\;c}
\tilde{\pi}^{cb}-\tilde{\pi}\tilde{\pi}^{ab}\right]+
\frac{\alpha}{2}\kappa^{-1/2}\sqrt{h}\left[h^{ab}\Box_{h}
N-{\cal D}_{h}^{a}{\cal D}_{h}^{b}N\right]+{\cal L}_{\vec{N}}
\tilde{\pi}^{ab}\nonumber
\end{eqnarray}
where both $R_{ab}$ and the covariant derivatives are built with
$h_{ab}$. 
From every solution to (\ref{42}) we can obtain a solution to the
EFE's for both Euclidean and Lorentzian gravity in the following way.
Choose
$\kappa$ and define new lapse and shift (${\cal N}$, and
${\cal N}^{a}$
respectively)  such that the Eqs. (\ref{44})  have exactly the form
of the equations derived from the Hamiltonian constraints (\ref{33})
or (\ref{34}) but written in terms of $h_{ab}$, $\tilde{\pi}^{ab}$,
${\cal N}$, and ${\cal N}^{a}$. In particular, let us take (\ref{43})
with  $\kappa=\frac{1}{2}\sqrt{|\alpha(\alpha+2\beta)|}$ and define
\begin{eqnarray}
& &{\cal N}^{a}=N^{a}\label{45}\\
& &{\cal N}=\frac{2N\kappa^{3/2}}{|\alpha+2\beta|}\nonumber
\end{eqnarray}
Note that, in general, the $q_{ab}$ from which $h_{ab}$
is built depends on $\alpha$ and $\beta$ in a non-trivial way. 
From these objects we can build the 4-dimensional metric
$g_{ab}^{Eins}$ solution to the EFE's for both Euclidean and
Lorentzian signatures if
we know $q_{ab}$, $N$, and $N^{a}$ (that we can get, for
example, by solving the field equations derived from (\ref{16})
after a certain foliation has been introduced). Although it is
possible to give a concrete prescription to build
$g_{ab}^{Eins}$ in terms of ${\cal N}$, ${\cal N}^a$, $h_{ab}$
and the foliation given by $\Phi$ it is better to do this
directly in four dimensions. This is the scope of the next
section.

\section{The Field Equations}

In this section we return to the four dimensional form of the action
(\ref{16}) and derive the field equations for a fixed $\Phi$. The 
solutions to them are always 4-metrics with Euclidean
signature. These metrics are taken as auxiliary fields from which we
can build solutions to the EFE's. We show how this is done
in this section and discuss some specific solutions in the next. In
order to show all the dependence on $g_{ab}$ of the action (\ref{16}),
and in order to simplify the field equations it is convenient to
rewrite it as 
\begin{equation}
S[g_{ab};\alpha,\beta]=\int_{{\cal M}} d^4x\sqrt{g}
\left[(\alpha+\beta)
R^{ab}\frac{\partial_a\Phi \partial_b\Phi}{g^{cd}\partial_c\Phi
\partial_d\Phi}-\frac{1}{2}\alpha R\right]
\label{38}
\end{equation}
Notice that this action reduces to the usual Euclidean 
Einstein-Hilbert action if $\alpha+\beta=0$, so the field equations 
must reduce exactly to the Euclidean EFE's in this case. Varying
(\ref{38}) with respect to $g_{ab}$ we get
\begin{eqnarray}
& &
\hspace{-1cm}\frac{\alpha}{2}G^{ab}+(\alpha+\beta)\left\{
\left[\frac{1}{2}g^{ab}+
\eta^a\eta^b\right]R_{cd}\eta^c\eta^d+\right.\nonumber\\
& &
\hspace{-1cm}+\left.\nabla_{c}\nabla^{(a}\left(\eta^{b)}
\eta^c\right)-\frac{1}{2}g^{ab}\nabla_c\nabla_d(\eta^c\eta^d)-2
\eta_d\eta^{(a}R^{b)d}-\frac{1}{2}\Box(\eta^a\eta^b)\right\}=0
\label{39}
\end{eqnarray}
where $\eta_a$ is given by (\ref{17}). According to the discussion in
the previous sections, for any given $\eta_a$ the solutions
to these equations give rise to solutions
to the EFE's either for Euclidean or Lorentzian general
relativity. For $\alpha\neq 0$ and
$\alpha+2\beta\neq 0$ all the solutions to (\ref{39}) can be
interpreted in this way and no new solutions are
introduced in the theory. Notice that, with the exception of
$\alpha+\beta=0$, for which (\ref{39}) actually reduces to the
Euclidean EFE's, the solutions to the previous equations (even if the
values of $\alpha$ and $\beta$ correspond to Euclidean gravity)
are not solutions to the EFE's themselves; we need
a prescription to build them from the solutions
to (\ref{39}) and the scalar field $\Phi$. From the argument in
section  III, and especially from (\ref{43}) and (\ref{45}) we find
out that $g^{Eins}_{ab}$ can be written in terms of $g_{ab}$ and
$\Phi$ as
\begin{equation}
g^{Eins}_{ab}=\frac{1}{2}\sqrt{|\alpha(\alpha+2\beta)|}\left
(g_{ab}-2\frac{\alpha+\beta}{\alpha+2\beta}\eta_{a}\eta_{b}\right)
\label{f1}
\end{equation}
In order to show that this is the case we will compute ${\cal N}$
and $h_{ab}$; the lapse and
the 3-metric defined by $g^{Eins}_{ab}$ and the foliation, and write
them in terms of $N$ and $q_{ab}$ (defined by $g_{ab}$ and $\Phi$). To
this end we need
\begin{eqnarray}
& & g_{Eins}^{ab}=\frac{2}{\sqrt{|\alpha(\alpha+2\beta)|}}\left(
g^{ab}-2\frac{\alpha+\beta}{\alpha}\eta^{a}\eta^{b}\right)\label{f2}\\
& &
\eta_{a}^{Eins}=\frac{|\alpha|}{2^{1/2}|\alpha(\alpha+2\beta)|^{1/4}}
\;\eta_{a}\label{f2a}
\end{eqnarray}
where $g_{Eins}^{ab}\eta_{a}^{Eins}\eta_{b}^{Eins}=-{\rm sgn}
[\alpha(\alpha+2\beta)]\equiv\zeta$,\footnote {$\zeta=+1$ 
and $\zeta=-1$ for Lorentzian and  Euclidean signatures respectively.}
Taking into account that 
\begin{equation}
{\cal N}=\frac{1}{|g^{ab}_{Eins}\partial_{a}\Phi\partial_{b}
\Phi|^{1/2}}
\label{f3}
\end{equation}
we obtain
\begin{equation}
\frac{{\cal N}}{N}=
\frac{(g^{ab}\partial_{a}\Phi\partial_{b}\Phi)^{1/2}}
{|g^{ab}_{Eins}\partial_{a}\Phi\partial_{b}\Phi|^{1/2}}=\frac{1}
{|g^{ab}_{Eins}\eta_{a}\eta_{b}|^{1/2}}=
\frac{|\alpha|}{2^{1/2}|\alpha(\alpha+2\beta)|^{1/4}}
\label{f4}
\end{equation}
which coincides with the result given by (\ref{45}). In the same
fashion we find 
\begin{equation}
h_{ab}=g_{Eins}^{ab}+\zeta\eta_{a}^{Eins}\eta_{b}^{Eins}=
\frac{1}{2}\sqrt{|\alpha(\alpha+2\beta)|}(g_{ab}-\eta_{a}\eta_{b})=
\frac{1}{2}\sqrt{|\alpha(\alpha+2\beta)|}\;q_{ab}
\label{f5}
\end{equation}
Notice that $g^{Eins}_{ab}$ will
depend, in general, on $\alpha$ and $\beta$, both through the
explicitly parameter dependent factors and the
$(\alpha,\beta)$-dependence of $g_{ab}$
and $\eta_{a}$. 

In view of (\ref{f1}) there is an alternative way to understand what
we are doing. Let us {\it define}
\begin{equation}
{\hat g}_{ab}\equiv\frac{1}{2}\sqrt{|\alpha(\alpha+2\beta)|}\left
(g_{ab}-2\frac{\alpha+\beta}{\alpha+2\beta}\eta_{a}\eta_{b}\right)
\label{f6}
\end{equation}
and compute
\begin{equation}
{\hat S}[g_{ab};\alpha,\beta]=-{\rm sgn}[\alpha]\int_{{\cal M}}d^4x\;
\sqrt{|\hat g|}R[\hat g_{ab}]
\label{f7}
\end{equation}
where $\hat g\equiv\det\hat g_{ab}$. Throwing away surface
terms we get, precisely, (\ref{38}) if $\alpha(\alpha+2\beta)\neq 0$,
so we conclude that our action (\ref{16}) reduces to the
Einstein-Hilbert action for metrics with either signature and
(\ref{39}) are the Einstein field equations. Notice, however, that if
$\alpha(\alpha+2\beta)=0$ then (\ref{f2}) is not defined because
$\hat g=0$. In this case the action (\ref{38}) and the field
equations (\ref{39}) provide a generalization of general relativity
for degenerate metrics. The degrees of freedom of this degenerate
theory are contained in the auxiliary Euclidean metric $g_{ab}$ even
if $\alpha=0$ although in this case we do not have the possibility of
interpreting them as describing a space-time metric (it would be zero,
according to (\ref{f6})).

Once we know that our action is the
Einstein-Hilbert action we can couple matter very easily, just by
adding the usual matter terms written with the help of
$\hat g_{ab}$. We will not discuss this issue further here.

In the canonical framework used in sections II and III we derived the
field equations for a fixed $\Phi$ and argued that the non-covariance
of the action should play no role as we are free to choose any
foliation when performing the Legendre transform from the
Lagrangian to the Hamiltonian formulation. Now to is clear why this is
so: we are just using the Einstein-Hilbert action written in the form
given by (\ref{f7}).

Let us consider now the case in which $\Phi$ is taken as a dynamical
field (and thus we have a covariant action). As we showed above, the
action (\ref{16}) is really the Einstein-Hilbert action for the
metric $\hat g_{ab}$ defined by (\ref{f6}). If the scalar field is
dynamical, i.e. not fixed, (\ref{16}) has a new gauge symmetry: the
invariance under local transformations of $\Phi$ that do not change
the combination of $g_{ab}$ and $\Phi$ defined by (\ref{f6}) (modulo
four-dimensional diffeomorphisms). The choice of a particular $\Phi$
should be considered as a gauge fixing of this new symmetry; given
two solutions to the field equations (\ref{39})
$(g_{ab}^{(1)},\Phi^{(1)})$ and $(g_{ab}^{(2)},\Phi^{(2)})$ it is
always possible, if $\alpha\neq 0$, to use the new gauge freedom in
order to have $\Phi^{(1)}=\Phi^{(2)}$, i.e. refer them to the same
scalar field. In this sense the particular choice of $\Phi$ is
irrelevant. It is also obvious that the additional
field equation obtained by varying the action with respect to $\Phi$
is redundant;  i.e. identically satisfied for any $g_{ab}$ and
$\Phi$ that solve (\ref{39}) because by varying $\Phi$ we only
generate a particular variation in ${\hat g}_{ab}$.

\section{Some Solutions to the New Field Equations}
Equation (\ref{39}) looks, arguably, more complicated than the usual
Einstein field equations, nevertheless some simple solutions to it
can be found, and equation (\ref{f1}) can be used to obtain some
familiar space-time metrics. We will concentrate on obtaining a
solution to (\ref{39}) that describes both Euclidean and Lorentzian
Schwarzschild black holes. Let us consider
\begin{eqnarray}
& & g_{ab}={\rm Diag}\left[1-\frac{2M}{r},(1-\frac{2M}{r})^{-1},
                           r^2,r^2 \sin^2\theta\right]\label{49a}\\
& & \eta_{a}=(1-\frac{2M}{r})^{1/2}[1,0,0,0]\label {49b}
\end{eqnarray}
Notice that the previous $\eta_{a}$ is obtained from $\Phi=x^0$ and 
has unit length. As it is well known the Ricci tensor computed from
the previous $\hat{g}_{ab}$ is zero so it is enough to show that
$\hat{g}_{ab}$ and $\eta_{a}$ satisfy
\begin{equation}
\nabla_{c}\nabla^{(a}\left(\eta^{b)}
\eta^c\right)-\frac{1}{2}g^{ab}\nabla_c\nabla_d(\eta^c\eta^d)
-\frac{1}{2}\Box(\eta^a\eta^b)=0
\label{50}
\end{equation}
A straightforward computation shows that this is indeed the case.
Plugging (\ref{49a},\ref{49b}) in (\ref{f1}) we get the following
solution to the EFE's
\begin{equation}
g_{ab}^{Eins}=\frac{1}{2}\sqrt{|\alpha(\alpha+2\beta)|}\;
{\rm Diag}\left[-\frac{\zeta|\alpha|}{|\alpha+2\beta|}
(1-\frac{2M}{r}),
(1-\frac{2M}{r})^{-1},r^2,r^2 \sin^2\theta\right]
\label{52}
\end{equation}
By suitably rescaling the $x^0$ and $r$ coordinates we can write
(\ref{52}) as the Schwarzschild solution with mass
$\bar{M}=2^{-1/2}|\alpha(\alpha+2\beta)|^{1/4}M$
and either Euclidean or Lorentzian signature. Hence we see how the
Schwarzschild solution is recovered in our formalism. Some features
of this solution are worth of discussion at this point. First of all
we notice that, in this case, $g_{ab}$ is independent of $\alpha$ and
$\beta$. This is not expected to be a feature of general solutions to
equation (\ref{39})
but only of those already satisfying $R_{ab}=0$. Second, it is
worthwhile pointing out that (\ref{52}) is not well defined for
$\alpha+2\beta=0$ and is zero for $\alpha=0$, but (\ref{49a}) is
always well defined, so there are solutions to the field equations in
the $\alpha=0$ case that do not admit a space-time interpretation.
Their degrees of freedom are contained in the auxiliary object
$g_{ab}$. Third, the
solution to (\ref{39}) describes a family of Schwarzschild black
holes with a mass that depends on $\alpha$ and $\beta$, and both
signatures.

\section{Comments and conclusions}

The main result presented in this paper is the introduction of a
modified action principle for general relativity that has the merit
of describing space-times of Lorentzian or Euclidean signature by
introducing some free parameters in the action. The solutions to the
field equations can be interpreted as families of solutions to the
EFE's whose signature can be chosen at will; some members of these
families are Lorentzian and some others are Euclidean. The problem
of finding a Euclidean solution associated to a certain
Lorentzian one can be solved in this framework by taking members
of a certain two parametric family of solutions to the new field
equations and choosing two of them with different signatures. At
least in some simple cases, like Schwarzschild, one gets the same
result as applying the usual Wick rotation (obtained by
introducing a purely imaginary time coordinate). The action used
in the paper can be useful in order to obtain generalizations of
general relativity and offers some interesting possibilities.
For example, the fact that (\ref{16}) is written in terms of a
Euclidean signature metric allows us to use compact internal
symmetry groups if we want to use a tetrad formalism. In this
respect it should be pointed out that the Ashtekar formalism for
general relativity \cite{Ash} can be derived from the, so
called,  Samuel-Jacobson-Smolin \cite{SJS} action that is
written in terms of a tetrad and a self-dual $SO(3,1)$
connection. The fact that
the internal group is $SO(3,1)$ has the consequence of requiring
the use of complex fields thus complicating the, otherwise,
very elegant formalism. Although there are some actions that
lead to real Ashtekar-like formulations \cite{Hol,fer} it
is interesting to explore alternative formulations in the spirit
of the present paper. 

Another interesting consequence of the analysis presented here 
is the realization of the fact that one can describe
a covariant theory with a non-covariant action, just notice that
for a fixed $\Phi$  the action (\ref{16}) may be written in the
obviously non-covariant form
\begin{equation}
S=\int_{{\cal M}} d^4x
\left[(\alpha+\beta)\frac{R^{00}}{g^{00}}-\frac{1}{2}\alpha R\right]
\label{54}
\end{equation}
where $\Phi=x^0$. Of course the fact that the field equations of
a covariant theory (or, rather, the solutions to them) can be
derived from a non-covariant action should not be considered as
very surprising. A finite function, for example, does not have
to be periodic in order to have periodic extrema; just consider
$f(x)=e^{-x}\sin^2x$. Not even the field equations have to be
periodic but only have periodic solutions as it happens in this
case. We briefly discuss now some applications of this
formalism.

i) {\it The problem of time}: The problem of understanding the
meaning of time in general relativity is a difficult one. For
example, one can try to find some function of the configuration
variable (a 3-metric in the ADM formalism or the Ashtekar
connection) such that evolution can be defined or described with
respect to it. In the asymptotically flat case, where the ADM
energy is different from zero, one can think of time as the
object canonically conjugate to it. In a rather loose sense it
can be said that this time is some kind of boundary condition
for certain operators that must be supplied by hand \cite{RT}.
The scalar field used in this paper has a natural interpretation
as time, not only because it provides slicings of the space-time
and is used to define "evolution",
but also because it gives us the time direction that we need in
order to get Lorentzian metrics from the auxiliary Euclidean
metric used in (\ref{f1}). In the asymptotically flat case a
function of $\Phi$ might be interpreted as the time
canonically conjugate to the ADM energy. This issue will be
considered in future work.
  
ii) {\it Quantum gravity}: As suggested in the introduction,
having the possibility of associating to any Lorentzian
solution to the EFE's a Euclidean one can be used
to derive some interesting properties of black holes, for
example the thermal nature of the Hawking radiation in the
Schwarzschild case. If the new field equations can be solved for
families containing more general black hole solutions (with or
without matter) it would be possible to study if the arguments
valid for the Schwarzschild black hole still hold in more
general cases. In a different context, the possibility of writing
actions for general relativity with different kinds of fields
may be useful in order to find extensions of the theory that,
considered from a perturbative point of view, may behave better
that the usual Einstein-Hilbert action or its higher derivative
extensions. 

iii) {\it Signature change}: Signature change has become a
popular subject since the suggestion by Hawking that the present
Lorentzian regime of gravity could have derived from a Euclidean
space-time via quantum tunneling. The issue of finding solutions
to the EFE's with signature change and, especially, the joining
conditions at the hypersurface of signature change has received
lots of attention in recent years \cite{Emb}. If the parameters
of the action (\ref{16}) can be provided with some dynamics the
metric (\ref{f1}) can change signature while still being a
solution to the EFE's. The issue of junction conditions may be
looked at not
in the solutions to the EFE's but in the auxiliary Euclidean
metric. Notice that in the Schwarzschild case considered above
the auxiliary metric is independent of the parameters and so,
even if $\alpha$ and $\beta$ ``evolve" in such a way that
$g^{Eins}_{ab}$ changes signature, $g_{ab}$ remains perfectly
regular. We do not know if this is true in more general
situations. In our opinion this issue deserves a closer
look and will be explored in the future.

The main open question that remains to be answered is the issue
of the regularity (analyticity?) of the solutions to the new
field equations in the parameters $\alpha$ and $\beta$. In
particular it would be desirable to see if the following statements 
are true:

i) If $g_{ab}^{(1)} (\alpha,\beta)$ and $g_{ab}^{(2)}
(\alpha,\beta)$ are solutions to (\ref{39}) corresponding to the same
$\Phi$, and $\alpha_0$, $\beta_0$,  $\alpha_1$, $\beta_1$ exist such
that $g_{ab}^{(1)} (\alpha_0 \beta_0)=g_{ab}^{(2)}(\alpha_1\beta_1)$
(modulo four-dimensional diffeomorphisms) then for any $\alpha$ and
$\beta$ it is possible to find $\alpha^{\prime}$ and $\beta^{\prime}$
such that $g_{ab}^{(1)} (\alpha,\beta)=g_{ab}^{(2)}(\alpha^{\prime}
\beta^{\prime})$ (modulo four-dimensional diffeomorphisms). In other
words; if the $g_{ab}^{(1)} (\alpha,\beta)$ and $g_{ab}^{(2)}
(\alpha,\beta)$ share an element then they are the same family of
solutions to (\ref{39}).

ii) Every solution to (\ref{39}) is sufficiently regular (for
example, analytic) in the parameters $\alpha$ and $\beta$.

This issues will be explored in future work.

{\bf Acknowledgments}
I want to thank my friends and colleagues for many useful
discussions, especially to Francisco Alonso, Juan Le\'on,
Alvaro Llorente, Peter Peld\'an, Juan P\'erez Mercader and
Alfredo Tiemblo. I am especially grateful to Carsten Gundlach
for his very enlightening comments and proof reading of the
manuscript and Tevian Dray for drawing my attention to some
references about signature change.

\newpage
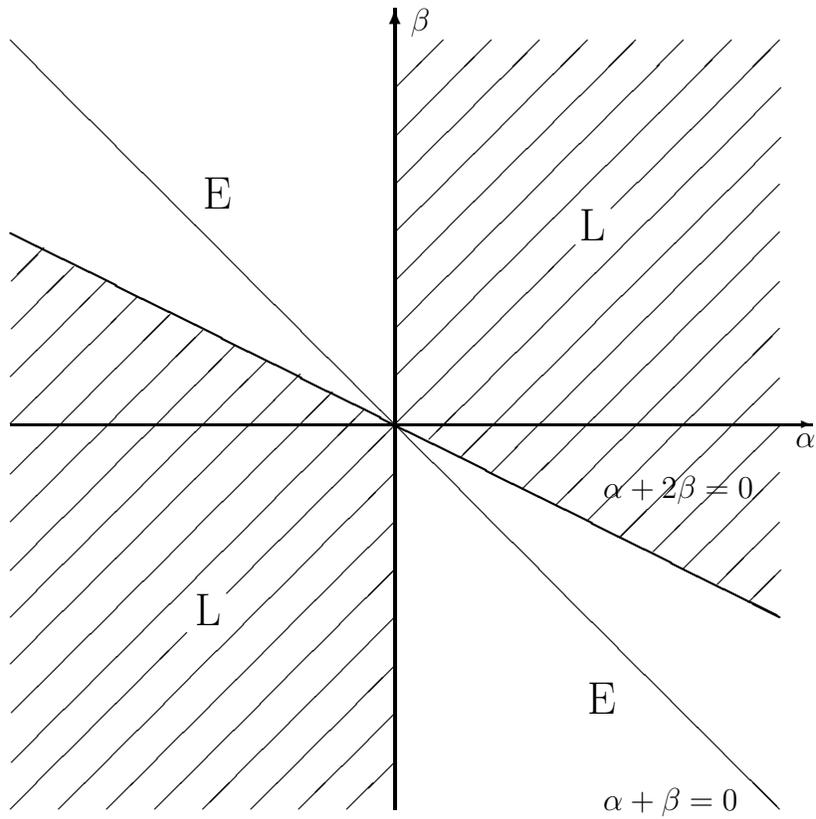
\begin{figure}[h]
\setlength{\unitlength}{0.00083300in}%

\begin{picture}(4831,6431)(-300,-6780)
\thicklines
\put(3301,-5761){\vector( 0, 1){5000}}
\thinlines
\put(901,-3361){\vector( 1, 0){5000}}
\thicklines
\put(901,-2161){\line( 2,-1){4800}}
\thinlines
\put(901,-961){\line( 1,-1){4800}}
\put(3301,-1261){\line( 1, 1){300}}
\put(3301,-1561){\line( 1, 1){600}}
\put(3301,-1861){\line( 1, 1){900}}
\put(3301,-2161){\line( 1, 1){1200}}
\put(3301,-2461){\line( 1, 1){1500}}
\put(3301,-2761){\line( 1, 1){1800}}
\put(3301,-3061){\line( 1, 1){2100}}
\put(3301,-3361){\line( 1, 1){1125}}
\put(4651,-2011){\line( 1, 1){1050}}
\put(1201,-5761){\line( 1, 1){2100}}
\put(3301,-3961){\line(-1,-1){1800}}
\put(1801,-5761){\line( 1, 1){1500}}
\put(3301,-4561){\line(-1,-1){1200}}
\put(2401,-5761){\line( 1, 1){900}}
\put(3301,-5161){\line(-1,-1){600}}
\put(3001,-5761){\line( 1, 1){300}}
\put(908,-5160){\line( 1, 1){2002.500}}
\put(901,-4853){\line( 1, 1){1807.500}}
\put(901,-5760){\line( 1, 1){1102}}
\put(2251,-4411){\line( 1, 1){1050}}
\put(908,-5468){\line( 1, 1){2205}}
\put(901,-4560){\line( 1, 1){1597}}
\put(901,-4268){\line( 1, 1){1403}}
\put(901,-3968){\line( 1, 1){1200}}
\put(901,-3668){\line( 1, 1){997.500}}
\put(901,-3353){\line( 1, 1){795}}
\put(901,-3060){\line( 1, 1){600}}
\put(908,-2760){\line( 1, 1){390}}
\put(908,-2460){\line( 1, 1){202.500}}
\put(5708,-1259){\line(-1,-1){2193.500}}
\put(5708,-1566){\line(-1,-1){1995}}
\put(5701,-1861){\line(-1,-1){1800}}
\put(4501,-3961){\line( 1, 1){1200}}
\put(5101,-4261){\line( 1, 1){600}}
\put(5701,-2159){\line(-1,-1){1605}}
\put(5701,-2466){\line(-1,-1){1395}}
\put(5701,-3066){\line(-1,-1){998}}
\put(5708,-3359){\line(-1,-1){795}}
\put(5701,-3966){\line(-1,-1){390}}
\put(5708,-4266){\line(-1,-1){202.500}}
\put(4451,-2211){\Large L}
\put(2051,-4611){\Large L}
\put(2101,-2011){\Large E}
\put(4501,-5161){\Large E}
\put(4600,-5763){$\alpha+\beta=0$}
\put(4600,-3820){$\alpha+2\beta=0$}
\put(5801,-3500){$\alpha$}
\put(3401,-900){$\beta$}
\end{picture}
\caption{Regions in the $(\alpha,\beta)$ plane corresponding to
Euclidean and Lorentzian signatures (E and L respectively). The
metrics in the line $\alpha+\beta=0$ are actual solutions to the
Euclidean EFE's. If $\alpha+2\beta=0$ the theory is singular and if
$\alpha=0$ it is not general relativity.}
\end{figure}

\end{document}